\documentclass[conference]{IEEEtran}
\IEEEoverridecommandlockouts
\usepackage{cite}
\usepackage{amsmath,amssymb,amsfonts}
\usepackage{algorithmic}
\usepackage{graphicx}
\usepackage{textcomp}
\usepackage{xcolor}

\usepackage{ntheorem}
\theoremseparator{:}
\newtheorem{hyp}{Assumption}

\def\BibTeX{{\rm B\kern-.05em{\sc i\kern-.025em b}\kern-.08em
    T\kern-.1667em\lower.7ex\hbox{E}\kern-.125emX}}
\begin{document}

\title{Complementing Handcrafted Features with Raw Waveform Using a Light-weight Auxiliary Model  \\
}

\author{\IEEEauthorblockN{Zhongwei Teng, Quchen Fu, Jules White, Maria Powell, Douglas C. Schmidt}
}

\maketitle

\begin{abstract}
An emerging trend in audio processing is capturing low-level speech representations from raw waveforms. These representations have shown promising results on a variety of tasks, such as speech recognition and speech separation. 
Compared to handcrafted features, learning speech features via 
backpropagation provides the model greater flexibility in how it represents data for different tasks theoretically.
However, results from empirical study shows that, in some tasks, such as voice spoof detection, handcrafted features are more competitive than learned features.
Instead of evaluating handcrafted features and raw waveforms independently, this paper proposes an Auxiliary Rawnet model to complement handcrafted features with features learned from raw waveforms. A key benefit of the approach is that it can improve accuracy at a relatively low computational cost. 
The proposed Auxiliary Rawnet model is tested using the ASVspoof 2019 dataset and the results from this dataset indicate that a light-weight waveform encoder can potentially boost the performance of handcrafted-features-based encoders in exchange for a small amount of additional computational work.

\end{abstract} 


\section{Introduction}

Fixed, handcrafted audio features, such as Mel-filter banks~\cite{fechner1966elements}, have shown great performance in capturing strong audio features in aspects of both auditory and machine learning~\cite{anden2014deep, zeghidour2021leaf}. 
However, since handcrafted features are often designed based on specific tasks, such as speech recognition, 
using these features to solve problems that they were not designed for may not be optimal.  
For example, Mel-filter banks~\cite{fechner1966elements} apply triangular filter banks on a Mel-scale to spectrograms calculated using short-term Fourier transform (STFT) to represent the non-linear perception of the human hearing. The Mel-scale is derived from a set of perception experiments on humans. 
As a result, Mel-filter banks are coarse-grained at high-frequencies since humans are less sensitive to high frequency sound. This loss of signal energy (information) in high frequencies may lead to poor performance on tasks that rely on information in these higher frequencies~\cite{zeghidour2021leaf}.

Extracting audio features with backpropagation provides an alternative way to represent raw waveforms by using deep neural networks to learn task-specific features. Task-specific features can be learned for many problems, such as voice recognition\cite{jung2018avoiding, jung2019rawnet} or automatic speaker verification (ASV)~\cite{liu2021tera}. Directly learning features from raw waveforms grants greater flexibility in handling unknown tasks and, thus, overcomes some of the challenges of handcrafted features, 
which may lose signal energy needed by a specific task.
Previous research indicates that representations learned from waveforms still have limitations on signal energy loss compared to the original raw signals they were learned from~\cite{anden2014deep}.
On certain tasks, such as Voice Spoof Detection, models based on handcrafted data still show much better performance than waveforms-based models~\cite{tak2021end,li2021replay}.

Instead of relying on raw waveforms independently, a potential option is to design a solution that can take advantage of 
both handcrafted and learned features.
For example, lost phase information in handcrafted features can be complemented by features learned from raw waveforms.
Even though there have been attempts to feed both handcrafted features and raw waveforms into networks for audio pattern recognition problems, limited research discusses the role of merging raw waveforms into arbitrary networks, as well as the trade-offs in model complexity of doing so.

In this paper, we propose the Auxiliary Rawnet (ARNet) architecture to combine learned features from raw waveforms with existing handcrafted features, by designing a lightweight auxiliary encoder.
The proposed model was tested on the ASV Spoof 2019 dataset~\cite{todisco2019asvspoof}, where the model needs defense against voice spoof attacks from various sources.
The model shows great promise in boosting the performance of single handcrafted-features-based networks that warrant further investigation on additional data sets and tasks.

The key contributions of this paper are as follows:
\begin{itemize}
\item We elaborate on the problem of concatenating raw waveforms and handcrafted features in the speech field and propose an assumption to solve this problem efficiently.
\item Based on our assumption, we introduce the Auxiliary Rawnet architecture that can be used to attach a lightweight auxiliary encoder to a model that relies on handcrafted features, so that raw waveform data can supplement the information in handcrafted features. 

\item We show results that indicate that, by introducing the auxiliary raw encoder, model performance is boosted on the ASV spoof 2019 dataset. 

\item We describe how our results show the potential of combining a light-weight waveform encoder with other encoders, providing an approach to balance the trade-off between performance and model complexity for models containing multiple encoders.
\end{itemize}

The remainder of this paper is organized as follows:
Section\ref{sec-related} discusses prior work in audio signal feature representation. 
Section\ref{sec-model} explains the problem analyzed in this paper and describes the Auxiliary Rawnet structure.
Section\ref{sec-experiment} introduces the experimental dataset and tasks used in this paper.
Experimental results are analyzed in Section\ref{sec-results}.
Section\ref{sec-conclusion} presents concluding remarks and lessons learned.

\section{Related Work}
\label{sec-related}
Prior work has shown how the "front-end" of models, which extract features from raw data, can be improved by using deep neural networks~\cite{anden2014deep, schneider2019wav2vec,jung2019rawnet, zeghidour2021leaf, ravanelli2018speaker} to directly learn features from raw signal data.
Directly applying standard CNNs to process raw waveforms~\cite{palaz2015convolutional} has shown promising results in speech recognition, spoofing detection, and speech separation.
Convolutions on time-domain raw waveforms can be explained as finite impulse response filter banks~\cite{anden2014deep}.
Structured filters are applied to optimize standard CNNs 
based on digital signal processing theory,
by initializing the first convolutional layer, which is believed to be the most important part, with known filter families~\cite{ravanelli2018speaker, noe2020cgcnn, balestriero2018spline}, so that a custom filter bank can be designed for a specific task.
Filter-based waveforms networks are emerging as excellent front-ends for many tasks~\cite{tak2021end,zeghidour2021leaf}.
However, a theoretical analysis from Joakim et al.~\cite{anden2014deep} has shown that signal energy loss is still inevitable for features extracted from raw waveforms by a CNN. Their results show extracted features can carry up to 94.5\% signal energy compared to the original waveforms.
On the other hand, empirical research also indicates that handcrafted features are still competitive in specific questions, such as speech commands~\cite{zeghidour2021leaf}, voice spoof detection~\cite{todisco2019asvspoof}, and instrument classification~\cite{zeghidour2021leaf}.

Even though there have been attempts to combine raw waveforms and handcrafted features in audio recognition~\cite{kong2020panns}, a general architecture for merging raw waveforms into networks that use handcrafted features, as well as the trade-offs in model complexity, have not been thoroughly investigated.
This paper considers the use of raw waveforms as a supplement to handcrafted features and investigates their potential to boost performance with little additional computational cost.

\begin{figure*}
    \centering
    \includegraphics[width=16cm,height=5cm]{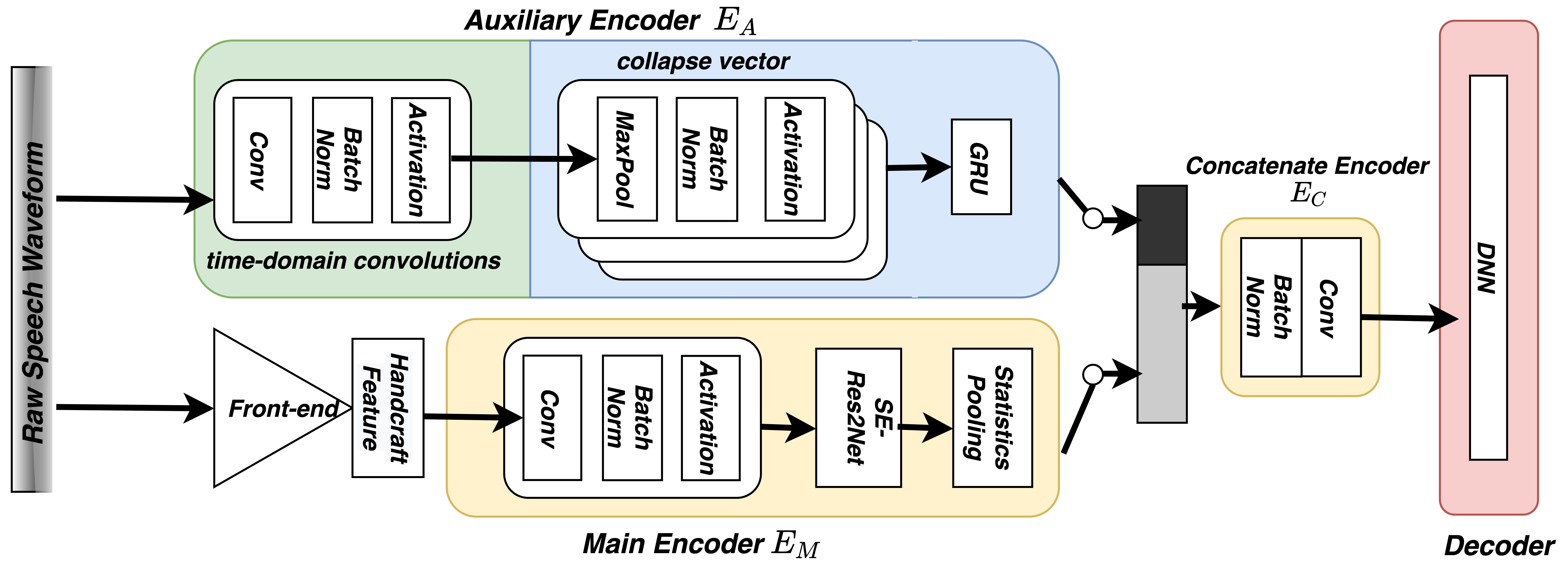}
    \caption{The ARNet Architecture. $E_A$ contains one strided CNN, 3 continuous max-pooling layers and a GRU.  A TDNN-based model is  illustrated here as an example of the $E_M$.} 
    \label{fig.structure}
\end{figure*}

\section{Auxiliary RawNet}
\label{sec-model}
\begin{figure}[hbtp]
     \centering
     \includegraphics[width=4cm,height=5cm]{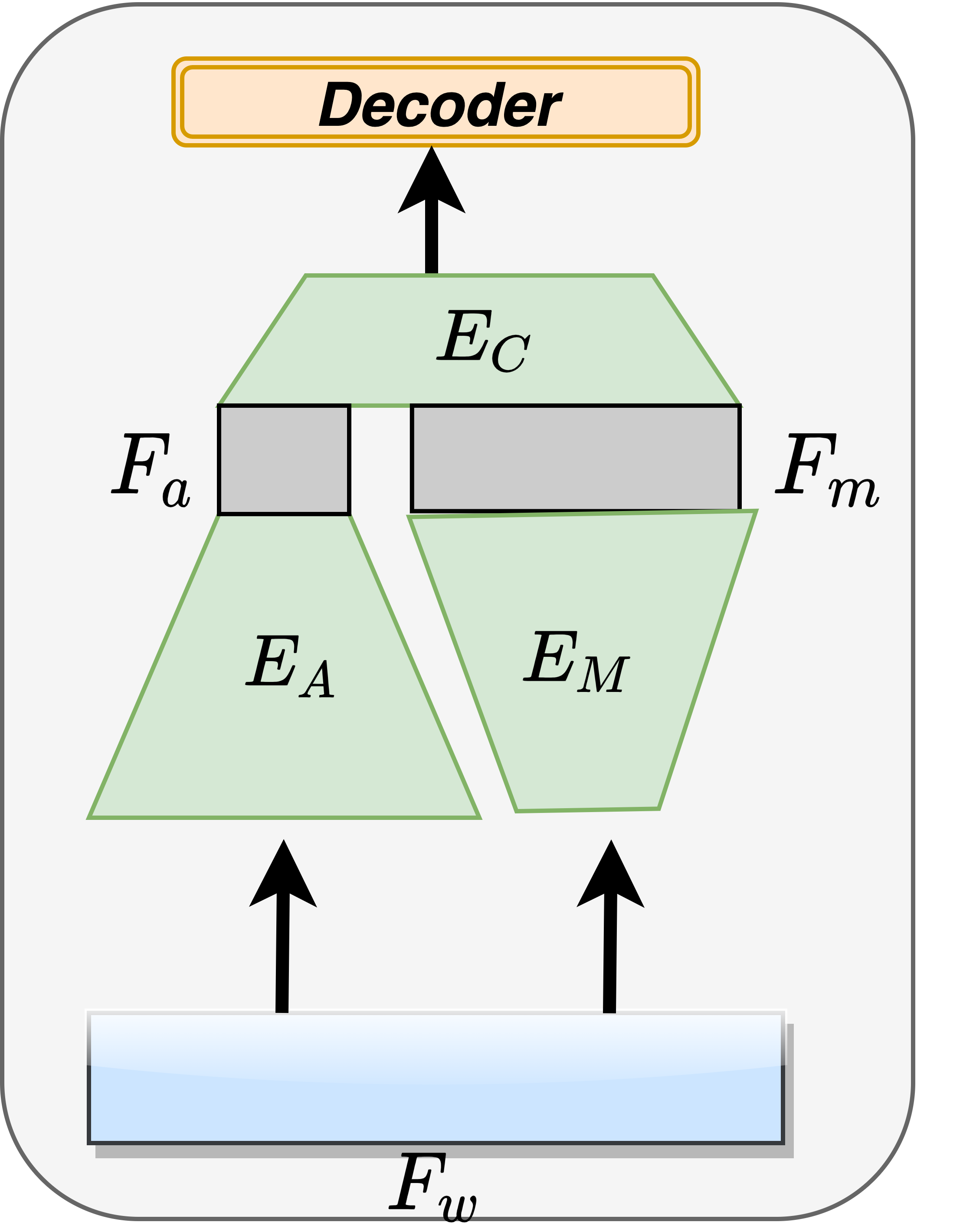}
     \caption{Overview of the ARawNet Architecture. The model consists of a Main Encoder($E_M$),  Auxiliary Encoder($E_A$), and Concatenate Encoder($E_C$). $E_A$ has a smaller bottleneck than $E_M$.}
     \label{fig-overview}
 \end{figure}

This section elaborates the research problem on combining raw waveforms and handcrafted features and explains the structure of the ARNet architecture.
The proposed network architecture applies a light-weight encoder to process raw waveforms with low computational cost as learned features, which are combined with existing speech classification models (Figure~\ref{fig.structure}).
To produce disentangled representations from different encoders, a narrow bottleneck is leveraged in the raw waveforms encoder without damaging the performance of the handcrafted encoder, as shown in Figure~\ref{fig-overview}.

\subsection{Problem Formulation}
\label{sec.problem}
Before introducing the ARNet architecture, we first formalize the problem that it is intended to solve.
Denote $F_{w}$ as features of a raw waveform, and $p$ as a problem to solve. We assume there is a constructive function $f$, which can map $F_{p_{mag}}$, $F_{p_{phase}}$ and $S_{p_{noise}}$ into $F_{w}$, as described in Equation~\ref{e_1}, where $F_{p_{mag}}$ is the ideal magnitude information needed to solve $p$,  $F_{p_{phase}}$ is the ideal phase information needed to solve $p$, and $S_{p_{noise}}$ are signals with
limited contribution to solving $p$ (e.g., background noise). 
\begin{equation} 
     \label{e_1}
     F_{w} = f(F_{p_{mag}}, F_{p_{phase}}, S_{p_{noise}})
\end{equation}

Empirical studies~\cite{zeghidour2021leaf} have shown the ability of handcrafted features to represent the strongest audio features for a variety of problems.
Based on our assumption, the calculation of handcrafted features can be denoted as a  mapping function $g$, which can retrieve approximations of $F_{p_{mag}}$ or $F_{p_{phase}}$.
For example, mel-spectragrams can be described by the following equation:

\begin{equation} 
     \label{e_2}
     F_{p_{mag}} \approx F_{mel} =  g_{mel}( |STFT(F_{w})|^2 )) 
\end{equation}

When concatenating 
raw waveform data and handcrafted features to enhance model performance, our work is essentially to find a function, $h$, so that the total loss of $g(F_w)$ and $h(F_w)$ is smaller than a single $g(F_w)$. 
In other words, we want to find representations closer to the ideal solution $F_{p_{mag}} +  F_{p_{phase}}$, as describe in Equation~\ref{e_5}.
\begin{equation} 
     \label{e_5}
     concat(g(F_{w}), h(F_{w})) \approx F_{p_{mag}} +  F_{p_{phase}} > g(F_{w})
\end{equation}

However, it is not clear how $g(F_{w})$ interacts with $h(F_{w})$. 
Inspired by observations from results regarding  $g(F_{w})$ and $h(F_{w})$ on various tasks~\cite{zeghidour2021leaf, tak2021end}, we make the following assumption about combining learned features and handcrafted features:

\begin{hyp}[A\ref{hyp:1}] \label{hyp:1}
If a handcrafted feature, $g(F_{w})$ shows strong results solving problem $p$, then there exists a   
$h(F_{w})$ with size less than $N$ in $concat(g(F_{w}), h(F_{w}))$ that will enhance overall performance.
In other words, $h(F_{w})$ can be an auxiliary component of $g(F_{w})$ to improve performance with a bounded cost. 
\end{hyp}

\subsection{The Auxiliary RawNet Structure}
Based on the assumptions presented in section~\ref{sec.problem}, we propose the ARNet architecture. 
An overview of the ARNet architecture is shown in Figure~\ref{fig.structure}. $E_A$, which processes the raw waveform, has a smaller bottleneck than $E_M$ which processes handcrafted audio features, to make the raw waveforms play a supplementary role and bound the computational cost (e.g., bound $N$). 

\textbf{The Encoders.}
There are 3 encoders in the ARNet: the Main Encoder($E_M$),  Auxiliary Encoder($E_A$), and Concatenate Encoder($E_C$). 
$E_M$ denotes the main encoder, whose inputs are the original handcrafted features that have shown good performance in solving the target problem.
$E_A$ is the encoder used to encode the raw waveforms in a light-weight way to compress $F_{w}$ into $F_{a}$, where $F_{a}$ are the features extracted by the auxiliary encoder.
$F_{a}$ and $F_{m}$ (hand crafted features from the main encoder) are then concatenated in channels and further encoded by $E_C$. 

Figure~\ref{fig.structure} shows details of the encoders used in our experiments on the ASVspoof 2019 dataset. 
We select the 
strided convolutional layer\cite{jung2019rawnet} as the first layer to directly process the raw waveforms.
However, unlike previous raw waveforms networks, which include multiple CNN blocks with large kernels,  the strided convolutional layer is only followed by 3 continuous pooling blocks to collapse vectors and remove any frame variance without further convolution.
A GRU is used to encode frame-level features into utterance-level embeddings by keeping output vectors from the last time step. 

The main encoder keeps layers before the statistical pooling layer, which will output utterance-level embeddings.
Based on our assumption~\ref{hyp:1}, we chose a narrow bottleneck for $E_A$. The dimension of the utterance-level embedding from $E_A$ is designed to be smaller than the output dimension from $E_M$.
In the end, $E_C$ only contains a single Conv1d to encode concatenated results from $E_A$ and $E_M$.

The full architecture and model hyper-parameters are explained in Table~\ref{table-parameters}.

\begin{table}[thpb]
  \centering
  \begin{tabular}{c|c}
  \hline
     Encoders  &  Blocks   \\ 
    \hline
    Auxiliary Encoder & Conv(3,3,128)  \\
         & BN\&LeakyReLu  \\ 
         & MaxPooling  \\
         & BN\&LeakyReLu  \\ 
         & GRU(512) \\ \hline
    Concatenate Encoder & BN  \\
    & Conv1D(1,1,256)  \\\hline
  \end{tabular}
    \vspace{2pt}
  \caption{The architecture of Auxiliary Encoder and Concatenate Encoder.} 
  \label{table-parameters}
\end{table}

\textbf{The Decoder.} In our problem, the decoder is a linear classifier layer that decodes embeddings from $E_C$ to target classification.

\subsection{Why does a light-weight encoded raw waveforms augment handcrafted features?}

(1) Compared to the current filter-based architectures as discussed in Section~\ref{sec-related}, we chose the strided convolutional receptive field, which is a standard CNN, as the first layer to process the raw waveforms.
The strided convolutional layer consists of a set of time-domain convolutions, where all parameters(CNN kernel), are learned from the data.
Calculation of the first CNN layer can be described as the following Equation~\cite{ravanelli2018speaker}, where x[n] is raw waveforms, h[n] is the filter and y[n] is filtered output:
\begin{equation} 
     \label{e_conv}
     y[n] = x[n] * h[n] = \sum_{0}^{L-1} x[l] \cdot  h[n-l]
\end{equation}

As discussed in  Section~\ref{sec.problem}, concatenating $g(F_{w})$ and $h(F_{w})$ requires each encoder to have different attention to features in the raw waveforms so that they can complement each other.
The standard convolutional layer with small kernels gives the $E_A$ the least information about the signal processing mechanisms in $g(F_{w})$, and thus potentially grants it the most flexibility to extract features, which do not overlap with  $g(F_{w})$.

(2) In contrast to previous waveform-based networks~\cite{jung2019rawnet, tak2021end}, the CNN blocks used in between the strided convolution layer and the GRU are completely removed, and only 3 continuous max-pooling layers with batch normalization are kept to collapse frame-level features step-by-step.

\begin{figure}[hbtp]
\centering
\includegraphics[width=8cm,height=5cm]{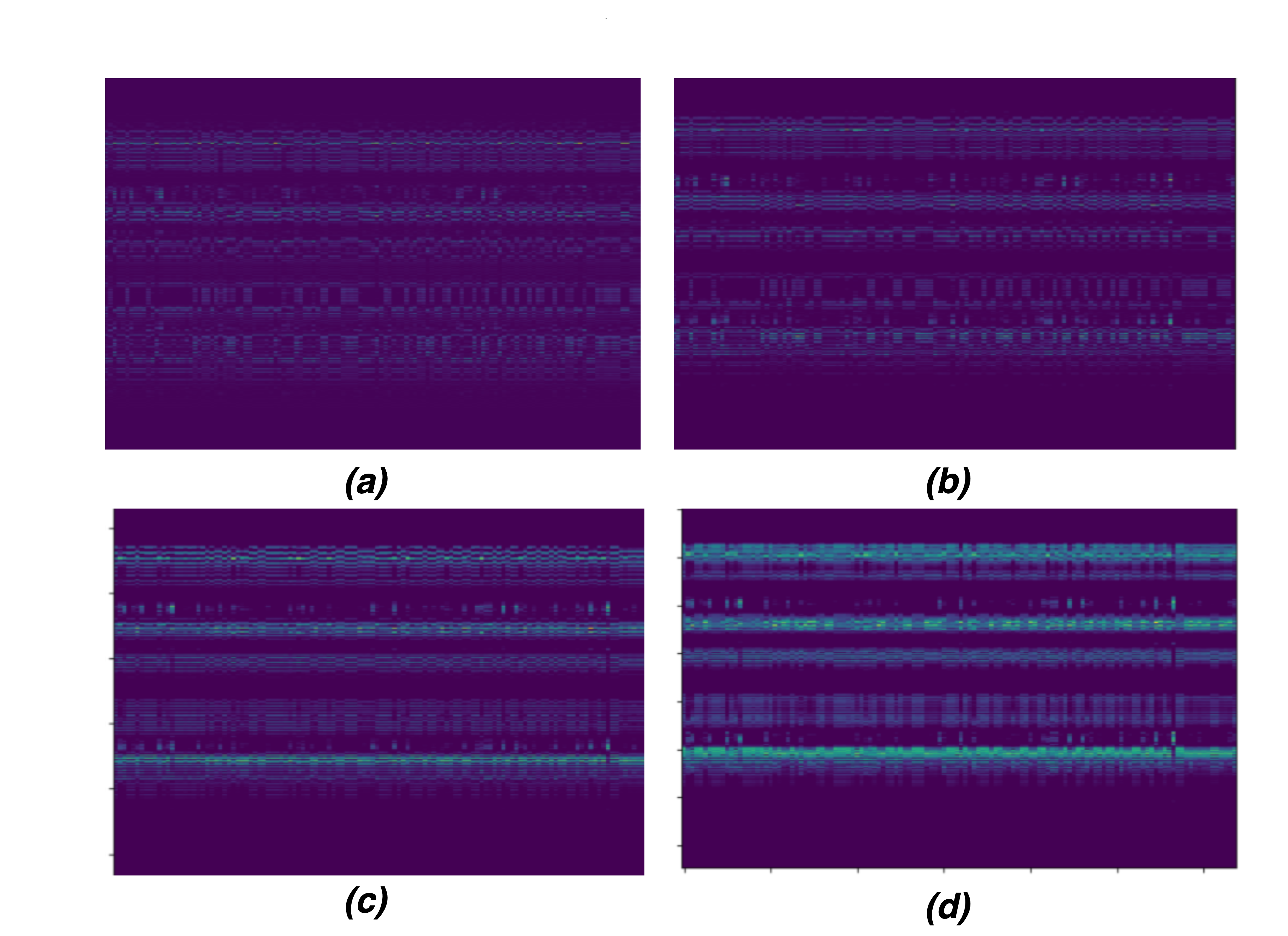}
\caption{Outputs visualization of the strided convolution layer and pooling layers. Outputs after 3 pooling layers(d) shows signification pattern information.}
\label{fig.pooling-vis}
\end{figure}

The first convolutional layer is considered the most critical part in processing raw waveforms. In deep networks it is also the  most vulnerable to problems, such as vanishing gradients, without initializing filters~\cite{ravanelli2018speaker}.
However, based on our assumption~\ref{hyp:1}, only significant frame-level features need to be kept, indicating networks without deep CNN blocks can be used for $E_A$.
Max pooling layers are used to collapse vectors and find significant pattern information that can be visualized after 3 pooling layers, as shown in Figure~\ref{fig.pooling-vis}. 

(3) We test our assumption~\ref{hyp:1} based on the Theorem~\cite{qian2020unsupervised} from speech conversion problems, that if information bottlenecks between different encoders are precisely set, the model will decompose and produce disentangled representations of input speech signals.
In our model, this Theorem can be described by the following equation:
\begin{equation} 
     \label{e_8}
     E_M(F_{w}) =  g(F_{w}),  E_A(F_{w}) = h(F_{w}) 
\end{equation}

Thus, a narrow bottleneck is designed for $E_A$, which means the dimension of  utterance-level embeddings $dim_{E_A}$  is much smaller than $dim_{E_M}$.

\section{Experimental Setup}
\label{sec-experiment}

\subsection{Experimental Dataset}
The  ASVspoof 2019 logical access (LA) dataset was developed to improve research on the growing threat of voice spoofing attacks on automated speech verification  systems~\cite{todisco2019asvspoof}.
This dataset contains human-recorded audios and spoof audios generated from 19 sources (A01 - A19), including speech synthesis, voice conversion, and hybrid algorithms. 
50,224 records in the training and development data  consist of spoof attacks generated by A01-A06. 
Another 71,237 spoof audio files in the evaluation data are generated by A07-A19, which are unpredictable spoofing attacks for CounterMeasure (CM) systems. 

We chose the ASVspoof 2019 LA dataset to validate the performance of our proposed model since:
\begin{itemize}

    \item The performance of handcrafted features is limited by the difference in spoofing sources between the training and evaluation data. 
    Spoofing types are highly unpredictable while the performance of CM systems relies on known spoofing attacks in training data and shows worse performance on unknown spoof attacks. 

    \item Current results on the ASVspoof 2019 challenge~\cite{todisco2019asvspoof,tak2021end} indicate that correct handcrafted features still provide the most competitive results from a single model compared raw waveforms approaches.
    
    \item Although pooling results of 19 spoof attacks is not satisfying, the waveforms-based network outperforms on the infamous A17 attacks~\cite{tak2021end}.

\end{itemize}

\subsection{Evaluation Metrics}
Two metrics are used to evaluate the ASVspoof 2019 LA dataset including \textit{min t-DCF} as the primary metric and $Equal Error Rate(EER)$ as a secondary metric, as described in~\cite{todisco2019asvspoof}: 

\subsubsection{\textit{min t-DCF}}
The Tandem Detection Cost Function (t-DCF)~\cite{kinnunen2018t} extends the conventional Detection Cost Function (DCF) in voice verification systems for spoofing attacks. 
The t-DCF measures the overall effect of  CM systems combined with existing ASV systems. The CM system acts as a gateway for the ASV system and this metric measures the overlapping of the two, a smaller value indicates better protection against spoofing. 

\subsubsection{$EER$}
EER indicates the threshold of a CM system where the false positive and false negative rates are equal each to other.

\subsection{Baseline Setup}
Our experiments include one handcrafted feature-based system and one raw waveforms-based system respectively:
\begin{itemize}
    \item \textbf{Res2net Architecture}. The Res2net architecture~\cite{li2021replay} is the state-of-the-art single system in the ASVspoof 2019 challenge, which tested the performance of 3 handcrafted features: log power magnitude spectrogram (Spec), linear frequency cepstral coefficients (LFCC), and constant-Q transform (CQT).
    \item \textbf{RawNet2}. The RawNet2~\cite{tak2021end} is the first anti-spoofing model, which only relies on the raw waveforms as input. It shows good performance on the A17 attack.
\end{itemize}

\begin{table}[thpb]
  \centering
  \begin{tabular}{cccccc}
  \hline
     & Front-end & Main Encoder & $E_A$ & EER & min-tDCF \\ 
    \hline
    \cite{li2021replay} & Spec & Res2Net\cite{li2021replay} & - & 8.783 & 0.2237 \\
    & LFCC &  & - & 2.869 & 0.0786 \\
    & CQT &  & - & 2.502 & 0.0743 \\
    \hline
    \cite{tak2021end} & Raw waveforms & Rawnet2\cite{tak2021end} & - & 5.13 & 0.1175 \\
    \hline
    \hline
    \hline
    Ours & Mel-Spectrogram & XVector & \checkmark & \textbf{1.32} & 0.03894 \\
     & &  & - & 2.39320 & 0.06875 \\
    \hline
    Ours & Mel-Spectrogram & ECAPA-TDNN	 & \checkmark & \textbf{1.39} & 0.04316 \\   
    &  &  & - & 2.11 & 0.06425\\ 
    \hline
    Ours & CQT & XVector & \checkmark & \textbf{1.74} & 0.05194 \\ 
    & &  & - & 3.39875 & 0.09510  \\ 
    \hline
    Ours & CQT & ECAPA-TDNN	 & \checkmark & \textbf{1.11} &   0.03645 \\
    & &. & - & 1.72667 & 0.05077 \\
    \hline
  \end{tabular}
    \vspace{2pt}
  \caption{Overall results of spoof detection systems to the ASVspoof 2019 dataset} 
  \label{table-model-results}
\end{table}

\section{Results and Analysis}
\label{sec-results}
Table~\ref{table-model-results} shows the experimental results of the ARNet on the ASVSpoof 2019 dataset. 
Results demonstrate the effectiveness of adding a light-weight auxiliary encoder to the main encoder.
Two  handcrafted features, Mel-spectrogram and CQT~\cite{schorkhuber2010constant}, as well as two state-of-the-art models in the speaker verification problem (XVector~\cite{snyder2018x, speechbrain} and ECAPA-TDNN~\cite{desplanques2020ecapa, speechbrain}) 
are selected as main encoders in the ARNet architecture.
Without modifying the hyper-parameters in the main encoder, we add the auxiliary encoder, as described in Table~\ref{table-parameters}, in the network to evaluate our assumption.
Overall, by introducing the auxiliary encoder, both $EER$ and $min-tDCF$ are reduced by \~ 50\% in all combinations of front-end and main encoders.
Specifically, CQT/ECAPA-TDNN with auxiliar encoder reaches the best performance on $EER$ of 1.11\% and $min-tDCF$ of 0.0364.

Table~\ref{table-model-size} compares  the number of trainable parameters and model complexity, multiply-and-accumulates (MACs) in our experiments. 
Compared to encoding handcrafted features (Res2Net), directly encoding raw waveforms (Rawnet2) increases model size and complexity by 2400\% and 600\%.
On the other hand, our auxiliary waveforms encoder only takes up 1.15M trainable parameters, which is a 19\% increase in ECAPA-TDNN and the model complexity increases from 2.36 GMac to 3.19 GMac.
In other words, the performance of our model increases by 28.2\% with increments of 35.1\% MACs. 

\begin{table}[thpb]
  \centering
  \begin{tabular}{|c|c|c|c|}
  \hline
     Main Encoder & Auxiliary Encoder & Parameters & MACs \\ 
    \hline
     Rawnet2 & - & 25.43 M & 7.61 GMac \\
     Res2Net & - & 0.92 M & 1.11 GMac \\
     XVector & \checkmark & 5.81 M  & 2.71 GMac \\
     XVector & - & 4.66M & 1.88 GMac  \\
     ECAPA-TDNN	 & \checkmark & 7.18 M & 3.19 GMac \\
     ECAPA-TDNN	 & - & 6.03M &  2.36 GMac \\
    \hline
  \end{tabular}
    \vspace{2pt}
  \caption{Comparison of model complexity (MACs) of various spoof detection systems} 
  \label{table-model-size}
\end{table}

\section{Conclusion and Future Work}
\label{sec-conclusion}

This paper discussed the problem of combining learned features and handcrafted featured in the audio field.
Based on our assumption that hand-crafted features and raw waveforms may complement each other without sacrificing model complexity, we proposed ARNet, which includes both hand-crafted features and raw waveforms as inputs. We tested 2 hand-crafted features (Mel-spectrogram and CQT) and 2 state-of-the-art models (XVector and ECAPA-TDNN) as the main encoder with our Auxiliary Encoder. 
Experiment results show raw waveforms has a general complementing ability to handcrafted features in the ASVspoof 2019 dataset.

\bibliographystyle{IEEEtran} 
\bibliography{main}

\begin{thebibliography}{10}
\providecommand{\url}[1]{#1}
\csname url@samestyle\endcsname
\providecommand{\newblock}{\relax}
\providecommand{\bibinfo}[2]{#2}
\providecommand{\BIBentrySTDinterwordspacing}{\spaceskip=0pt\relax}
\providecommand{\BIBentryALTinterwordstretchfactor}{4}
\providecommand{\BIBentryALTinterwordspacing}{\spaceskip=\fontdimen2\font plus
\BIBentryALTinterwordstretchfactor\fontdimen3\font minus
  \fontdimen4\font\relax}
\providecommand{\BIBforeignlanguage}[2]{{%
\expandafter\ifx\csname l@#1\endcsname\relax
\typeout{** WARNING: IEEEtran.bst: No hyphenation pattern has been}%
\typeout{** loaded for the language `#1'. Using the pattern for}%
\typeout{** the default language instead.}%
\else
\language=\csname l@#1\endcsname
\fi
#2}}
\providecommand{\BIBdecl}{\relax}
\BIBdecl

\bibitem{fechner1966elements}
G.~Fechner, ``Elements of psychophysics. vol. i.'' 1966.

\bibitem{anden2014deep}
J.~And{\'e}n and S.~Mallat, ``Deep scattering spectrum,'' \emph{IEEE
  Transactions on Signal Processing}, vol.~62, no.~16, pp. 4114--4128, 2014.

\bibitem{zeghidour2021leaf}
N.~Zeghidour, O.~Teboul, F.~d.~C. Quitry, and M.~Tagliasacchi, ``Leaf: A
  learnable frontend for audio classification,'' \emph{arXiv preprint
  arXiv:2101.08596}, 2021.

\bibitem{jung2018avoiding}
J.-W. Jung, H.-S. Heo, I.-H. Yang, H.-J. Shim, and H.-J. Yu, ``Avoiding speaker
  overfitting in end-to-end dnns using raw waveform for text-independent
  speaker verification,'' \emph{extraction}, vol.~8, no.~12, pp. 23--24, 2018.

\bibitem{jung2019rawnet}
J.-w. Jung, H.-S. Heo, J.-h. Kim, H.-j. Shim, and H.-J. Yu, ``Rawnet: Advanced
  end-to-end deep neural network using raw waveforms for text-independent
  speaker verification,'' \emph{arXiv preprint arXiv:1904.08104}, 2019.

\bibitem{liu2021tera}
A.~T. Liu, S.-W. Li, and H.-y. Lee, ``Tera: Self-supervised learning of
  transformer encoder representation for speech,'' \emph{IEEE/ACM Transactions
  on Audio, Speech, and Language Processing}, 2021.

\bibitem{tak2021end}
H.~Tak, J.~Patino, M.~Todisco, A.~Nautsch, N.~Evans, and A.~Larcher,
  ``End-to-end anti-spoofing with rawnet2,'' in \emph{ICASSP 2021-2021 IEEE
  International Conference on Acoustics, Speech and Signal Processing
  (ICASSP)}.\hskip 1em plus 0.5em minus 0.4em\relax IEEE, 2021, pp. 6369--6373.

\bibitem{li2021replay}
X.~Li, N.~Li, C.~Weng, X.~Liu, D.~Su, D.~Yu, and H.~Meng, ``Replay and
  synthetic speech detection with res2net architecture,'' in \emph{ICASSP
  2021-2021 IEEE International Conference on Acoustics, Speech and Signal
  Processing (ICASSP)}.\hskip 1em plus 0.5em minus 0.4em\relax IEEE, 2021, pp.
  6354--6358.

\bibitem{todisco2019asvspoof}
M.~Todisco, X.~Wang, V.~Vestman, M.~Sahidullah, H.~Delgado, A.~Nautsch,
  J.~Yamagishi, N.~Evans, T.~Kinnunen, and K.~A. Lee, ``Asvspoof 2019: Future
  horizons in spoofed and fake audio detection,'' \emph{arXiv preprint
  arXiv:1904.05441}, 2019.

\bibitem{schneider2019wav2vec}
S.~Schneider, A.~Baevski, R.~Collobert, and M.~Auli, ``wav2vec: Unsupervised
  pre-training for speech recognition,'' \emph{arXiv preprint
  arXiv:1904.05862}, 2019.

\bibitem{ravanelli2018speaker}
M.~Ravanelli and Y.~Bengio, ``Speaker recognition from raw waveform with
  sincnet,'' in \emph{2018 IEEE Spoken Language Technology Workshop
  (SLT)}.\hskip 1em plus 0.5em minus 0.4em\relax IEEE, 2018, pp. 1021--1028.

\bibitem{palaz2015convolutional}
D.~Palaz, M.~M. Doss, and R.~Collobert, ``Convolutional neural networks-based
  continuous speech recognition using raw speech signal,'' in \emph{2015 IEEE
  International Conference on Acoustics, Speech and Signal Processing
  (ICASSP)}.\hskip 1em plus 0.5em minus 0.4em\relax IEEE, 2015, pp. 4295--4299.

\bibitem{noe2020cgcnn}
P.-G. No{\'e}, T.~Parcollet, and M.~Morchid, ``Cgcnn: Complex gabor
  convolutional neural network on raw speech,'' in \emph{ICASSP 2020-2020 IEEE
  International Conference on Acoustics, Speech and Signal Processing
  (ICASSP)}.\hskip 1em plus 0.5em minus 0.4em\relax IEEE, 2020, pp. 7724--7728.

\bibitem{balestriero2018spline}
R.~Balestriero, R.~Cosentino, H.~Glotin, and R.~Baraniuk, ``Spline filters for
  end-to-end deep learning,'' in \emph{International conference on machine
  learning}.\hskip 1em plus 0.5em minus 0.4em\relax PMLR, 2018, pp. 364--373.

\bibitem{kong2020panns}
Q.~Kong, Y.~Cao, T.~Iqbal, Y.~Wang, W.~Wang, and M.~D. Plumbley, ``Panns:
  Large-scale pretrained audio neural networks for audio pattern recognition,''
  \emph{IEEE/ACM Transactions on Audio, Speech, and Language Processing},
  vol.~28, pp. 2880--2894, 2020.

\bibitem{qian2020unsupervised}
K.~Qian, Y.~Zhang, S.~Chang, M.~Hasegawa-Johnson, and D.~Cox, ``Unsupervised
  speech decomposition via triple information bottleneck,'' in
  \emph{International Conference on Machine Learning}.\hskip 1em plus 0.5em
  minus 0.4em\relax PMLR, 2020, pp. 7836--7846.

\bibitem{kinnunen2018t}
T.~Kinnunen, K.~A. Lee, H.~Delgado, N.~Evans, M.~Todisco, M.~Sahidullah,
  J.~Yamagishi, and D.~A. Reynolds, ``t-dcf: a detection cost function for the
  tandem assessment of spoofing countermeasures and automatic speaker
  verification,'' \emph{arXiv preprint arXiv:1804.09618}, 2018.

\bibitem{schorkhuber2010constant}
C.~Sch{\"o}rkhuber and A.~Klapuri, ``Constant-q transform toolbox for music
  processing,'' in \emph{7th sound and music computing conference, Barcelona,
  Spain}, 2010, pp. 3--64.

\bibitem{snyder2018x}
D.~Snyder, D.~Garcia-Romero, G.~Sell, D.~Povey, and S.~Khudanpur, ``X-vectors:
  Robust dnn embeddings for speaker recognition,'' in \emph{2018 IEEE
  International Conference on Acoustics, Speech and Signal Processing
  (ICASSP)}.\hskip 1em plus 0.5em minus 0.4em\relax IEEE, 2018, pp. 5329--5333.

\bibitem{speechbrain}
M.~Ravanelli, T.~Parcollet, P.~Plantinga, A.~Rouhe, S.~Cornell, L.~Lugosch,
  C.~Subakan, N.~Dawalatabad, A.~Heba, J.~Zhong, J.-C. Chou, S.-L. Yeh, S.-W.
  Fu, C.-F. Liao, E.~Rastorgueva, F.~Grondin, W.~Aris, H.~Na, Y.~Gao, R.~D.
  Mori, and Y.~Bengio, ``{SpeechBrain}: A general-purpose speech toolkit,''
  2021, arXiv:2106.04624.

\bibitem{desplanques2020ecapa}
B.~Desplanques, J.~Thienpondt, and K.~Demuynck, ``Ecapa-tdnn: Emphasized
  channel attention, propagation and aggregation in tdnn based speaker
  verification,'' \emph{arXiv preprint arXiv:2005.07143}, 2020.

\end{thebibliography}

\end{document}